      [1999/12/01 v1.4c Il Nuovo Cimento]
\documentclass{cimento}

\usepackage{graphicx}
\usepackage{epsfig}
\usepackage{subfig}
\usepackage{fancyhdr}
\usepackage{empheq}
\usepackage{mathbbol}
\usepackage{wasysym}
\usepackage{pstricks}
\usepackage{color}
\usepackage{bm}
\usepackage{multirow}

\title{Indirect search of exotic mesons: $B\to J/\psi + Anything$}
\author{
C.~Sabelli\from{ins:z}
}
\instlist{
\inst{ins:z} Department of Physics Universit\`a di Roma `Sapienza' and INFN Roma, \\ Piazzale A.~Moro 2, Roma, I-00185, Italy}
\PACSes{\PACSit{14.40.Rt}{Exotic Mesons}
\PACSit{13.85.Ni}{Inclusive production with identified hadrons}
\PACSit{13.25.Hw}{Decays of bottom mesons}}

\begin{document}

\maketitle

\begin{abstract}
We reconsider the discrepancy between theory and data in the momentum distribution of slow $J/\psi$ in $B$ decays.
Beside an update of the standard color singlet and color octet QCD components, we include the contribution from $XYZ$ exotic mesons, 
and show that the residual discrepancy could be accommodated considering new $XYZ$ mesons still unobserved.
\end{abstract}

\begin{figure}[t]
\centering
\includegraphics[width=10.0truecm]{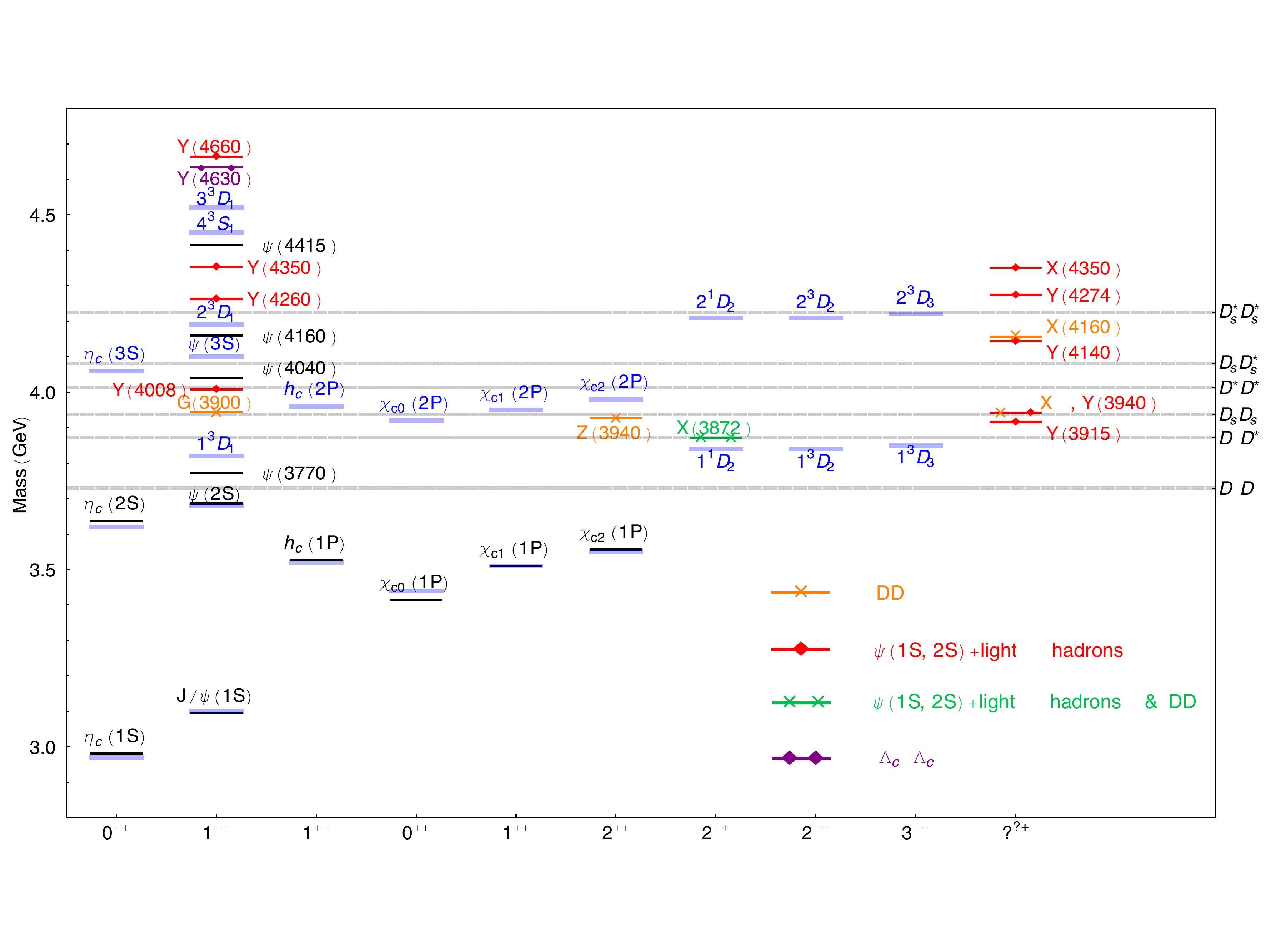}
\caption{The spectrum of $c\bar c$ states, predictions (solid thick blue lines) and experimental data (solid black lines), together with the known~\cite{Brambilla:2010cs} $XYZ$ mesons (colored lines with symbols),
which are classified according to their decay modes indicated in the legend.}
\label{fig:cspectrum}
\end{figure}

In the last ten years the search for excited charmonium and bottomonium states has revealed the existence of a number of resonances, named $XYZ$.
Even if containing $Q\bar Q$, $Q=c,b$, in their decay products, almost none of the $XYZ$ mesons can be interpreted as a standard $Q\bar Q$ structure,
showing production and decay rates in contrast with predictions from potential models.
In Fig.~\ref{fig:cspectrum} we show the known~\cite{Brambilla:2010cs} $XYZ$ mesons classified according to their decay modes and compared to the standard $c\bar c$ levels.
There are various phenomenological interpretations for the $XYZ$ mesons. Among the most explored possibilities:
({\it i}) {\it hadronic molecules}: bound states of two mesons interacting with each other via pion exchange;
({\it ii}) {\it tetraquarks}: compact clusters made up by a diquark, a $[qq]_{\bar{\bm{3}}_{\rm c}}$ state, and an antidiquark, a $[\bar q\bar q]_{\bm{3}_{\rm c}}$ state;
({\it iii}) {\it hybrids}: $q\bar q g$ aggregates;
({\it iv}) {\it hadrocharmonium}: a heavy quarkonium state $Q\bar Q$ embedded inside light hadronic matter.
Until now the discovery of the exotic mesons has been fairly accidental, it occurred when studying final states expected for the decays of higher standard $Q\bar Q$ states.
Here we consider the possibility to reveal the $XYZ$ not directly, but indirectly, taking into account processes in which they behave as intermediate states.
In particular we discuss the inclusive production of $J/\psi$ in $B$ decays and consider all the contributions separately following Ref.~\cite{Burns:2011rn}.

\begin{figure}
\begin{minipage}[t]{6.2truecm}
\centering
\includegraphics[width=6.0truecm]{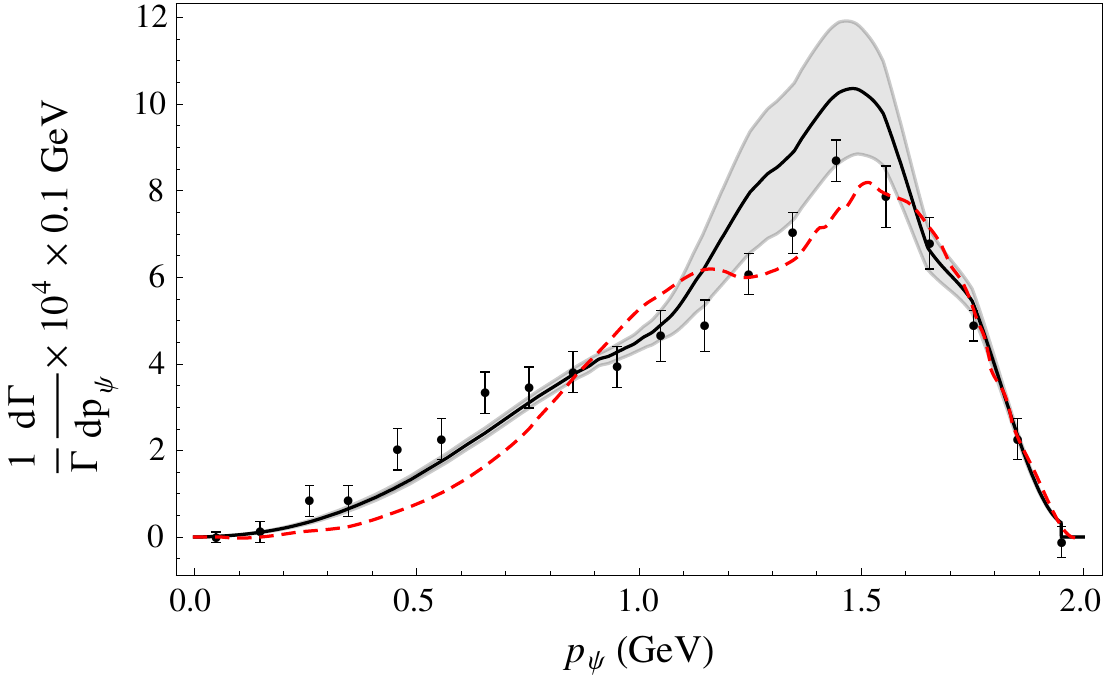}
\caption{Comparison between the old theoretical prediction~\cite{Aubert:2002hc} (red-dashed line) and the sum of the three contributions (black-solid line):
color singlet, color octet (with $\Lambda_{_{\rm QCD}}=500$~MeV and $p_{_F}=500$~MeV) and $XYZ$ mesons.
Data points from~\cite{Aubert:2002hc}.}
\label{fig:new_old}
\end{minipage}
\hspace{1truecm} 
\begin{minipage}[t]{6.2truecm} 
\centering
\includegraphics[width=6.0truecm]{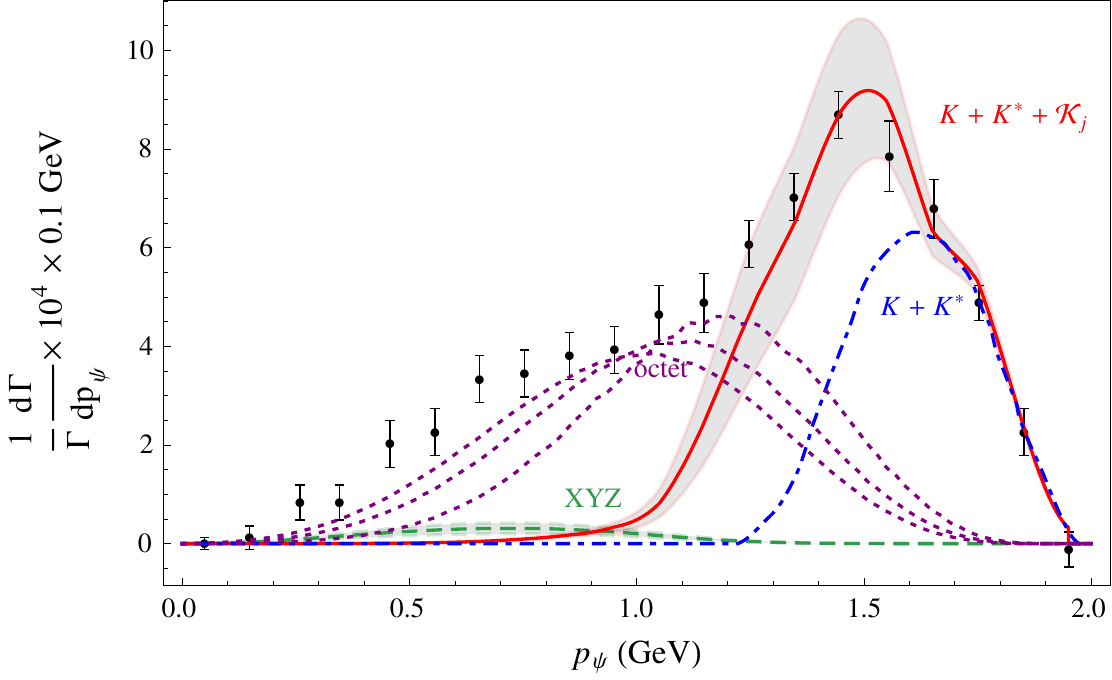}
\caption{Color singlet $B\to \mathcal{K}J/\psi$: blue dot-dashed represents $\mathcal{K}=K,K^*$~\cite{Aubert:2002hc}, red-solid includes also the heavier kaons~\cite{:2010if}.
Color octet: purple-dotted lines for three values of $\Lambda_{_{\rm QCD}}=300,500,800$~MeV and $p_{_F}=300$~MeV from right to left respectively.
$XYZ$: green-dashed line.}
\label{fig:diff2}
\end{minipage}
\end{figure}

{\bf \emph{The data}}. The BaBar collaboration~\cite{Aubert:2002hc} measured the decay momentum ($p_\psi$) distribution of the of the prompt $J/\psi$'s 
coming from $B$-mesons decays, Fig.~\ref{fig:new_old} (black-disks).

The theoretical description~\cite{Aubert:2002hc} in terms of two standard QCD components, reveals an excess of events in the low $p_\psi$ region
(red-dashed line in Fig.~\ref{fig:new_old}). This discrepancy has provoked a variety of phenomenological conjectures.

Beside the standard QCD components, on which we give an update based on the most recent data, we include here the contribution of the process: $B\to \mathcal{K}\mathcal{X}\to \mathcal{K}\; J/\psi\;+\;{\rm light}\; {\rm hadrons}$,
where $\mathcal{X}$ indicates one of the $XYZ$ mesons, and $\mathcal{K}$ a generic kaon.

{\bf \emph{Color singlet}}.
The high momentum side of the spectrum is filled by two-body decays of the kind $B\to \mathcal{K}\;J/\psi$,
where the $c\bar c$ pair is produced directly in color singlet configuration and thus hadronizes into $J/\psi$ without the emission of any gluon.
In Ref.~\cite{Aubert:2002hc} only $B\to K\;J/\psi$ and $B\to K^*\;J/\psi$ were included (blue dot-dashed line of Fig.~\ref{fig:diff2}).
Nevertheless a recent analysis by Belle~\cite{:2010if} on the $K\pi\pi$ invariant mass spectrum in the decays $B\to J/\psi\;K\pi\pi$ indicates
the existence of other two-body modes with heavier kaons, beside confirming the importance of the $K_1(1270)$ one.
From Belle measurements we extract the two-body branching ratios $\mathcal{B}(B\to \mathcal{K}J/\psi)$ with ${\mathcal K } = K_1(1270), K_1(1400), K^*(1410), K_2^*(1430), K_2(1600), K_2(1770), K_2(1980)$.
They are reported in Table I of Ref.~\cite{Burns:2011rn} and combined to give  the red-solid line of Fig.~\ref{fig:diff2}.

{\bf \emph{Color octet}}.
The low momentum region is associated to events where the $c\bar c$ pair is produced in color octet configuration, and thus emits soft gluons to fragment into $J/\psi$. 
As proposed in Ref.~\cite{Beneke:1999gq} the effect of the emitted gluons on the momentum distribution
can be modeled using a non relativistic shape function (with a characteristic energy scale of $m_c v^2\approx\Lambda_{_{\rm QCD}}$). 
Moreover the motion of the $b$ quark inside the $B$ meson can be described using the model given in Ref.~\cite{Altarelli:1982kh} with a characteristic momentum $p_{_F}$, the Fermi momentum.
The shape of the color octet component is ruled by the values of $(\Lambda_{_{\rm QCD}},p_{_F})$, 
but the absolute normalization, which depends on the non perturbative non-relativistic QCD (NRQCD) matrix elements, needs to be adjusted to data.

{\bf \emph{XYZ mesons}}.
Some of the $XYZ$ mesons have been observed in $B$-decays, produced in association with a scalar kaon with known branching ratios (see Table II of Ref.~\cite{Burns:2011rn}).
Among them $X(3872)$, $Y(3940)$, $Y(4140)$ and $Y(4260)$ decay into $J/\psi$ and light hadrons, and thus do contribute to the $p_\psi$ spectrum.
Besides the $B\to K\mathcal{X}$ modes, we decided to include also $B\to \mathcal{K}\mathcal{X}$ ones, where $\mathcal{K}$ are all the heavier kaons allowed by kinematics.
To determine the relative branching ratios, which are not measured, 
we use a simple scaling rule for the couplings deduced from a partial wave analysis. 
All the $XYZ$ here are considered to be $J=1$ resonances. 
The results are represented by the green-dashed line of Fig.~\ref{fig:diff2}. Even if very small, the contribution from exotic mesons peaks exactly in the region of the discrepancy between theory and experiment,
as first observed in Ref.~\cite{Bigi:2005fr}.

{\bf \emph{Results}}.
The complete distribution is obtained by summing up the color singlet, color octet and $XYZ$ components.
The values of $(\Lambda_{_{\rm QCD}},p_{_F})$ and of the absolute normalization of the octet curve are used as free parameters.
To best fit data (black-solid curve in Fig.~\ref{fig:new_old}) we choose the octet component with $\Lambda_{_{\rm QCD}}=500$~MeV and $p_{_F}=500$~MeV. 
Both these values are critically on the high sides of the allowed ranges,
$\Lambda_{_{\rm QCD}}\in [200,450]$~MeV and $p_{_F}\in [300,450]$~MeV. 
Yet the black-solid curve in Fig.~\ref{fig:new_old} represents a considerable improvement 
with respect to the old one (red-dashed). Relying on the validity of the NRQCD approach, our results indicates that 
the inclusion of new resonances of the $XYZ$ kind feeding the low $p_\psi$ region would effectively improve the agreement with data.

\bibliography{csabelli}

\end{document}